# Probabilistic representation and inverse design of metamaterials based on a deep generative model with semi-supervised learning strategy

*Wei Ma[1], Feng Cheng[2], Yihao Xu[1], Qinlong Wen[1] and Yongmin Liu[1,2,*]*

[1] Department of Mechanical and Industrial Engineering, Northeastern University,

Boston, Massachusetts 02115, United States

[2] Department of Electrical and Computer Engineering, Northeastern University,

Boston, Massachusetts 02115, United States

[*]E-mail: y.liu@northeastern.edu

**Abstract**

The research of metamaterials has achieved enormous success in the manipulation of light in an artificially prescribed manner using delicately designed sub-wavelength structures, so-called meta-atoms. Even though modern numerical methods allow to accurately calculate the optical response of complex structures, the inverse design of metamaterials, which requires to retrieve the optimal structure according to given requirements is still a challenging task due to the non-intuitive and non-unique relationship between physical structures and optical responses. To better unveil this implicit relationship and thus facilitate metamaterial design, we propose to represent metamaterials and model the inverse design problem in a probabilistically generative manner, believed to be the first attempt to do so. By employing an encoder-decoder configuration, our deep generative model compresses the meta-atom design and optical response into a latent space, where similar designs and similar optical responses are automatically clustered together. Therefore, by sampling in the latent space with certain prior distribution, the stochastic latent variables function as codes, from which the candidate designs are generated upon given requirements in a decoding process. With the latent variable as an effective representation of metamaterials, we can elegantly model the complex structure-performance relationship in an interpretable way, and solve the one-to-many mapping issue that is intractable in a deterministic model. Moreover, to alleviate the burden of numerical calculation when collecting data, we develop a semi-supervised learning strategy that allows our model to utilize unlabeled data in addition to labeled data during training, simultaneously optimizing the generative inverse design and deterministic forward prediction in an end-to-end manner. On a data-driven basis, the proposed deep generative model can serve as a comprehensive and efficient tool that accelerates the design, characterization and even new discovery in the research domain of metamaterials and photonics in general.

## Introduction

Artificially constructed metamaterials make it possible to design macroscopic optical responses from microscopic sub-wavelength structures that mimic atoms in ordinary materials [1-3]. Many exotic optical properties have been demonstrated experimentally based on various metamaterials, including negative refractive index [2], broadband chirality [4], strong nonlinearity[5], which lead to important applications, such as electromagnetic cloaking [6], perfect absorption [7], super lensing [8] and wavefront manipulation [9]. Although the degrees of freedom in the meta-atom design provides tremendous flexibility in engineering the optical properties of metamaterials, this bottom-up design strategy can hardly be generalized into concrete and practical guidelines. At present, the design process of metamaterials mostly relies on physics-inspired methods, resorting to human knowledge such as physical insights revealed by simplified analytical modeling, similar experience transferred from previous practice and intuition obtained by scientific reasoning. For example, many meta-atoms inherited traditional antenna design with geometries like rectangle [10], cross [11], bowtie [12], V-shape [13], H-shape [14], and so on, whose first-order response is approximated by electrical dipole resonance with relevant scaling effect [15]. Some other designs guided by physical intuition include ring-like structures that exhibit strong magnetic resonances induced by the incident magnetic field [6, 16, 17], dielectric building blocks that can induce both electric and magnetic resonances leading to better control of the phase of the scattered light [18-20], or the spectra line-shape tailoring by introducing coupling among different resonant modes [21-23].

Despite the exciting results obtained by these physics-inspired designs, this methodology basically relies on a trial-and-error process, usually involving numerical methods like Finite-Difference-Time-Domain (FDTD) or Finite Element Method (FEM) to iteratively solve Maxwell's equations. The low efficiency and thus limited exploration of the design varieties tend to easily omit the optimal solution. The inverse design approaches start from the opposite end, and try to optimize certain objective functions describing the desired performance [24, 25]. Common approaches for inverse problems include genetic algorithm [26], level set methods[27] and topology optimization[28], which, however, are still stochastic searching algorithms that are time-consuming and deteriorate rapidly as the design space grows. Different from numerical calculations, data-driven methods based on machine learning (ML) solve the optimization problem from statistical perspectives, so that the solution to optimize a target can be approximately generalized from numerous design examples. With the rapidly accumulated data and thus booming of deep learning (DL), the state of the art in many research domains, such as speech recognition[29], computer vision[30, 31], natural language processing [32] and decision making [33], has been pushed far beyond conventional methods. Deep neural networks simulate biological signal processing that allow computational models to learn multiple levels of abstract representations of data layer by layer [34], with superior advantages to discover intricate structures in large data sets by using the backpropagation algorithm in training. More recently as a powerful tool, DL has been applied to many other research fields such as material science [35, 36], chemistry [37-39], laser physics [40], particle physics [41], quantum mechanics [42, 43] and microscopy [44], showing great potential to circumvent the drawbacks of traditional methods in these areas.

Neural networks have been used to solve the design, optimization and prediction problems of electromagnetism in some early works [45-47], but the model capability and performance were limited, largely due to the simple model structure and the lack of data. More recent works sought to deal with the inverse design by DL under various scenarios, like plasmonic waveguide [48], optical power splitter [49], plasmonic metamaterials [50], chiral metamaterials [51] and nanophotonic particles [52]. Despite different design targets and network architectures, the common idea behind these works is to model the relationship between the design parameters and optical response as a bidirectional mapping, which is only able to deal with a few design parameters in a small range of applications. More fatally, treating metamaterial design as a one-to-one mapping as regression problem is inconsistent with the physical intuitions, since drastically different meta-atom structure can produce very similar optical response. To remedy this issue, tandem training strategy has been used to avoid instable training loss and force the inverse network to converge to one possible solution [53], but this operation sacrifices the design varieties and leads to limited generalization ability.

To mitigate all these challenges, here we propose a probabilistic graphic model as a comprehensive solution to metamaterial design problem. The metamaterial patterns are represented as two-dimensional images with the ability to describe any random geometries beyond fixed parameterization [54, 55]. Our model aims to follow the heuristic design procedure from human experience in the sense that similar geometries are clustered together as possible candidates, and similar optical responses are obtained by varying the design within each geometry group. To achieve this goal, for the first time, we introduce latent variables as a probabilistic representation of metamaterial design by incorporating a variational auto-encoder (VAE) structure in our model [56], which encodes the designed pattern together with the corresponding optical response into a latent space. By forcing a given prior distribution on the latent space, new designs can be reconstructed from the latent variables by sampling in the latent space, allowing many design varieties satisfying the same requirements. Unlike a very recently proposed generative model using generative adversarial network (GAN) [55] that needs a pre-trained simulator to guarantee the inverse design process, our model solves both the forward and inverse problem at the same time, and can be trained in an end-to-end manner. More notably, by introducing latent variable to encode metamaterial design, the proposed deep generative model offers interpretability and can utilize unlabeled data in a semi-supervised learning strategy to assist the construction of the latent distribution [57], which fully exploit human experience on possible metamaterial geometries while alleviating the intensive burden of numerical calculations to collect data. With the probabilistic representation capability, our model can produce much more diverse candidates for the inverse design, which, in conjunction with human knowledge, would open a new paradigm for discovering sophisticated metamaterials and photonic structures with prescribed, exotic optical properties and functionalities. In the meantime, the developed metamaterials and photonic platforms in turn may help to realize all-optical implementation of various machine learning and artificial intelligence techniques as demonstrated by recent exciting works [58, 59].

**Results and Discussions**

Figure 1(a) schematically shows the proposed framework to model metamaterials, which can simultaneously realize forward prediction from given designs and inverse retrieval of the possible designs from required optical responses. As shown in the left part of the figure, the meta-atom under investigation is a sandwiched structure working in the reflective configuration, which is composed of a continuous metallic ground plane, a dielectric spacer and a top metallic resonator. The thickness of the resonator and spacer is fixed at 50 and 100 nm, respectively, while the period of the unit cell is 2μm in both directions. We represent metamaterial design as a two-dimensional binary image of resolution 64×64, where 1 stands for metal and 0 stands for air. In this way, arbitrary design geometries can be properly represented, avoiding the limit of parameterization. Three basic geometry groups of cross, split ring and h-shape are selected as empirical designs for training. Since the continuous back reflector eliminates transmission, the optical response of the metamaterial is completely described by its reflection coefficients. Considering two different linear polarization conditions and optical reciprocity, we will focus on three unique reflection spectra, that is, $x$ polarization input with $x$ polarization output ($R_{xx}$), $y$ polarization input with $y$ polarization output ($R_{yy}$), and $x$ polarization input with $y$ polarization output ($R_{xy}$), which is identical to another cross-polarized reflection ($R_{yx}$). The reflection spectra of interest are set in the mid-infrared region from 40 THz to 100 THz, and discretized into 31 data points each with 2 THz interval. More details about data preparation can be found in the method section.

From physical insights, the forward prediction is a many-to-one mapping, since any meta-atom structure supporting the same resonant mode can produce the same optical response within a certain range of tolerance, which can be modeled deterministically as a regression problem. While the inverse design has to deal with the opposite, one-to-many mapping or structured output problem, allowing diverse predictions [60]. To solve this problem, we introduce a latent variable which, subject to a given prior distribution, encodes the meta-atom pattern together with its optical response. Therefore, the inverse design can be modeled as a generative process, by first sampling a latent variable from the latent space, and then reproducing the design from the sampled variable. This process introduces diversity due to the stochastic sampling step. The overall comprehensive problem including forward prediction, inverse generation, and latent space construction are solved jointly by the proposed deep generative model. We further illustrate the one-to-many mapping issue and the advantage of our model in Section 1 of the Supplementary Material.

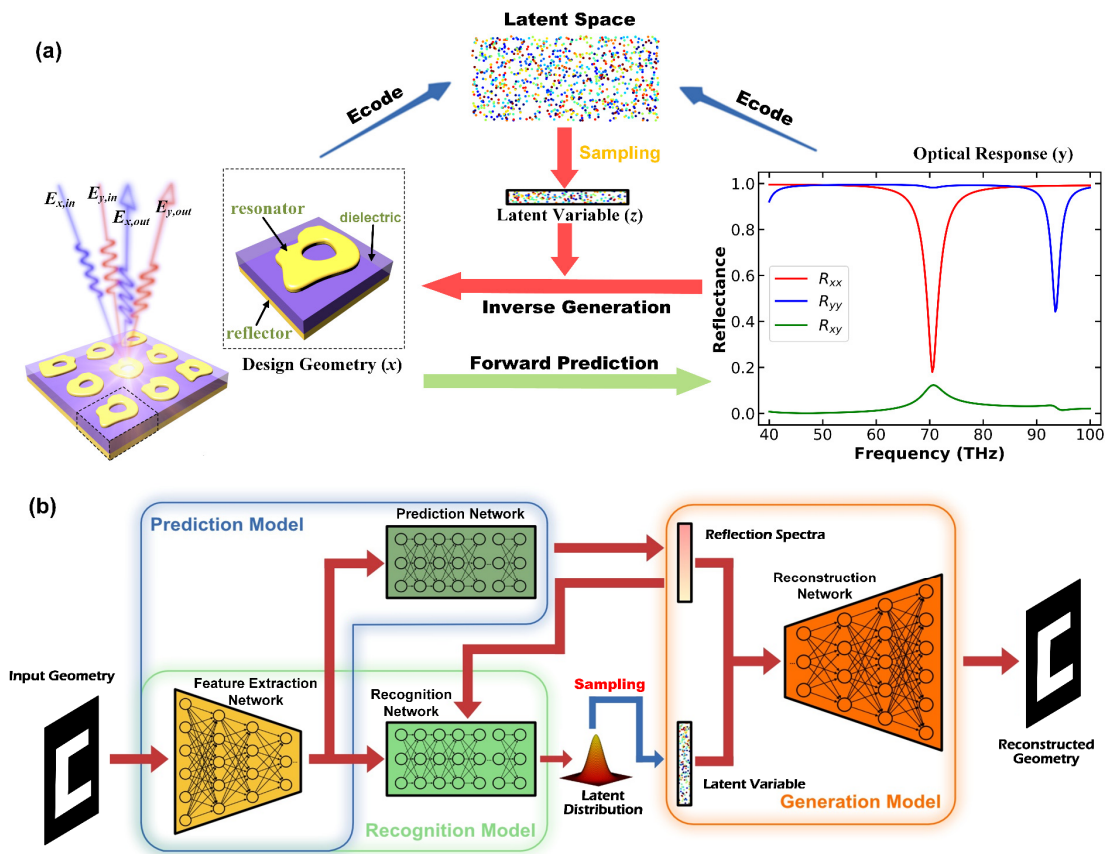

Figure 1: (a) The proposed deep generative model for metamaterial design and characterization. The metamaterial design and optical response are encoded into a latent space with a pre-defined prior distribution, from which the latent variables are sampled for inverse generation. The forward path is modeled as a deterministic prediction process. (b) Architecture of the proposed

deep generative model. Three sub-models, the recognition model, the prediction model and the generation model, constitute the complete architecture, which is implemented by four neural networks with deliberately designed structures for different purposes. The recognition model encodes the metamaterial pattern with its optical response into a low-dimensional latent space. The prediction model outputs a deterministic prediction of the optical response given the metamaterial design. The generation model accepts the optical response and the sampled latent variable to produce feasible metamaterial designs according to specific requirements.

**Mathematical formulation of the deep generative model**

As indicated in Figure 1(a), there are three types of variables in our deep generative model: input variable $x$ (geometric pattern of metamaterial structure), output variable $y$ (three distinct reflection spectra), and latent variable $z$ (compressed code of the design). The overall model builds up the probabilistic relationship among these variables, which accounts for different functionalities when applying the model to the metamaterial design. Shown as dashed boundaries in Figure 1(b), our deep generative model can be decomposed into three sub-models as follows:

**Recognition model:** The recognition model defines the posterior distribution of latent variable $z$ given input variable $x$ and the associated spectra $y$, $P_r(z|x, y)$, which models the encoding process of metamaterial structures into latent space.

**Prediction model:** The prediction model defines the conditional distribution of output variable $y$ given input variable $x$, $P_p(y|x)$, which models the forward prediction of reflection spectra of metamaterial with given geometric pattern.

**Generation model:** The generation model defines the generative distribution of geometric pattern $x$ given reflection spectra $y$ and latent variable $z$, $P_g(x|y, z)$, which models the inverse design process of metamaterial given required spectra.

Finally, the two major tasks of metamaterial design can be readily solved by this unified model. For the forward problem, only the prediction model is needed, which directly gives a deterministic prediction of reflection spectra given the input geometric pattern of the metamaterial. The inverse problem, due to its nature of one-to-many mapping, is treated as a generative process as follows: a set of latent variable $z$ is generated from the prior distribution $P_\theta(z)$, and the geometric data $x$ is generated by the generative distribution conditioned on latent variable $z$ and the given spectra $y$: $z \sim P_\theta(z)$, $x \sim P_\theta(x|y,z)$.

Although the prediction model can be optimized by conventional stochastic gradient descent with backpropagation algorithm, parameter estimation of the generation model is challenging due to intractable posterior inference [60]. However, by using the variational lower bound of the log-likelihood as a surrogate objective function, the model parameters can be efficiently estimated by stochastic gradient variational Bayes (SGVB) method [56]. The loss objective $L_l(x, y)$ for labeled data in the form of variational lower bound is written as:

$$\log p_\theta(x|y) \geq - KL[q_\emptyset(z|x,y)||p_\theta(z)] + \mathbb{E}_{q_\emptyset(z|x,y)}[\log p_\theta(x|y,z)] = -L_l(x,y) \quad (1)$$

In SGVB method, the proposal distribution $q_\emptyset(z|x,y)$ is introduced to approximate the true posterior, which is defined by the recognition model as $P_r(z|x,y)$. Assuming Gaussian latent variables, the first KL-divergence term of Equation 1 can be marginalized, while the second expectation term can be approximated by sampling from the recognition distribution $P_r(z|x,y)$. To utilize unlabeled data in the recognition model and generation model, the values of *y* obtained from prediction model $P_p(y|x)$ account for the evaluation of the generative loss similar to the supervised case. More detailed derivation of the variational lower bound and the semi-supervised training strategy can be found in Section 2 of the Supplementary Material.

**Implementation of the deep generative model**

In order to realize the deep generative model that provides a comprehensive solution to the metamaterial design and characterization, we implement the three probabilistic sub-models using four deep neural networks, each with deliberately designed structure for its specific function. Specifically, they are feature extraction network, prediction network, recognition network and reconstruction network as shown in Figure 1(b). We assume the latent variable is subject to multivariate Gaussian distribution and the dimension of the latent space is chosen to be 20 in our modeling. As shown in Figure 1(b), the metamaterial geometry is first fed into a feature extraction network, which is shared by both prediction model and recognition model. The feature extraction network comprises four convolutional layers with batch normalization (BN) followed by a fully connected layer, which essentially compresses the input two-dimensional data into a compact feature vector. The prediction network and recognition network are then built on top of the extracted features. For the prediction model, we use three parallel fully connected modules for the three distinct reflection spectra, that is, $R_{xx}$, $R_{yy}$ and $R_{xy}$, respectively. For the recognition model, two more fully connected layers are employed to convert the feature vector into the statistical parameters of the recognition distribution $P_r(z|x,y)$, that is the mean vector and the diagonal covariance matrix. The generation model is primarily composed of a reconstruction network, which accepts the three reflection spectra and the latent variable sampled from the recognition distribution, to reproduce the two-dimensional binary image representing the metamaterial design. The inverse functionality of the reconstruction network in comparison with the feature extraction network is implemented by transposed-convolutional layers that have the opposite effect of convolutions, gradually up-sampling the feature vector to a two-dimensional image.

The entire deep generative model is trained in an end-to-end manner with both labeled and unlabeled data. The total loss includes generation loss (Equation 1 and Equation S5) that accounts for latent space configuration and input image reconstruction, and prediction loss (Equation S6) to optimize forward prediction accuracy. We use Adam optimizer and the hyper-parameter $\alpha$ in Equation S7 is set to 2000. The batch size is 100 and the model was trained for 200 epochs when it converged. In each step, the labeled data are fed as complete pairs to the model, which contribute to both prediction loss and generation loss. The unlabeled data, which are only responsible to the generation loss in Equation S5, are paired with the corresponding predicted spectra to guarantee the data consistency during the entire end-to-end training. The details about network configuration and training process are described in Section 3 and Section 5 of the Supplementary Material.

**The latent space and pattern reconstruction**

In the proposed deep generative model, the recognition model and generation model together form a VAE structure. The VAE, conditioned on the corresponding reflection spectra, encodes the meta-atom pattern into latent space from which the original pattern can be reconstructed. During the training process, the VAE utilizes both labeled and unlabeled data, trying to reproduce input image with loss described as the expectation term in Equation 1. In Figure 2(a), the evolution of the reproduced images is illustrated at certain training steps. Given the input images on the leftmost column, our model gradually produces more and more accurate reconstruction of the inputs as the training proceeds. Finally, at the step of 300000 when training ends, the test input images are faithfully reconstructed with high fidelity, indicating an accurate distribution of meta-atom patterns described by the proposed model.

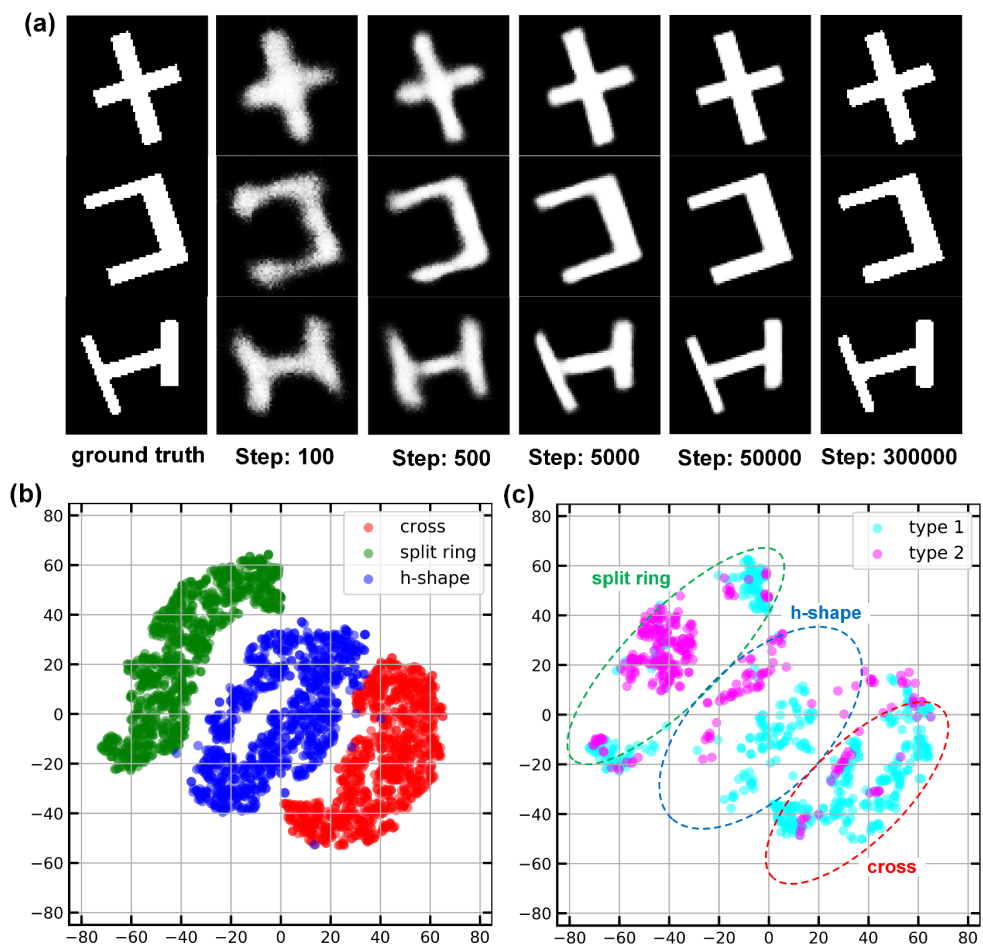

Figure 2: Reproduced geometry during training process after certain steps (a). Visualization of the latent space by reducing the dimension from 20 to 2 using t-SNE. The distribution of the meta-atom design highlighted by its shape (b) and by the two spectra types (c).

Before evaluating the performance of the proposed deep generative model, we first check the structure of the latent space where the meta-material design is encoded. Since the dimension of the latent space is 20, we use t-distributed stochastic neighbor embedding (t-SNE) method to reduce the dimension to 2 for visualization purpose. In Figure 2(b) we plot the 2D distribution of the encoded test data in three geometry groups. The three geometry groups are clearly separated into three clusters of cross, split ring and h-shape respectively. Without providing the corresponding labels for the shapes, our model automatically learns to distinguish different shapes through the encoding-decoding training iterations on both labeled and unlabeled data. As described by the KL-divergence term in Equation 1, we have applied a standard Gaussian prior distribution of the latent variables independent of the geometric patterns and reflection spectra, $p_\theta(z) \sim \mathcal{N}(0, I)$. The result in Figure 2(b) indicates that the latent distribution learned by our model nicely approximates the spherical Gaussian prior in the 20-dimensional latent space, while keeping the fine structure within the prior according to different design patterns.

On the other hand, by feeding both image features and reflection spectra into the recognition network (Figure 1(b)), the VAE structure actually models the distribution of design patterns conditioned on its optical response. Naturally, besides the shape information, the latent variables are expected to encode the information on optical response at the same time. Therefore, we proceed to check, within each geometry clusters, how their respective optical responses are distributed. Since the reflection spectra are continuous and thus cannot be divided into categories like design geometries, we choose to manually classify the optical response by certain criteria to see whether the latent space can distinguish different spectra line shape to some extent. Out of all 3000 test samples, we first pick the designs that have strong resonant responses, with the dips of spectra $R_{xx}$ and $R_{yy}$ smaller than 0.7 and the peak of spectrum $R_{xy}$ smaller than 0.3. Then the designs with the strongest resonance of $R_{xx}$ at a frequency smaller than 60 THz are defined as type 1 optical response, while those with the strongest resonance of $R_{xx}$ at a frequency larger than 60 THz are defined as type 2 optical response. Figure 2 (c) illustrates the distribution of the selected test data according to their different optical responses (different geometry groups are roughly indicated by the dashed ellipse). Not surprisingly, the two types of optical responses are separated into different clusters within each geometry group, especially for split ring and h-shape. In the split ring group, type 1 responses are clustered at two ends while type 2 responses are centered in the middle. In the h-shape group, the two types of response are also separated with a clear gap. Therefore, the latent variables learned by our model, with the dimension of only 20, not only encode the design patterns with the size of 64×64, but also reflect the associated optical response with the dimension of 93 in a compact but informative way.

**Model Evaluation**

The proposed deep generative model is a comprehensive system for metamaterial design, representation and characterization. As shown in Figure 3, we demonstrate the performance of our model by evaluating it on three samples randomly drawn from the three geometry groups in the test data set. The ground-truth geometry and reflection spectra obtained by numerical simulation are plotted as solid lines and insets in Figure 3(a), 3(d) and 3(g). The forward prediction performance is first evaluated by feeding the ground-truth geometry to the prediction model, and the output spectra, each discretized into 31 data points, are plotted as scattered hollow circles in the same figures. The excellent agreement between the predicted spectra by our model and the numerically simulated spectra clearly confirms that our model can function as an effective simulator for fast metamaterial characterization.

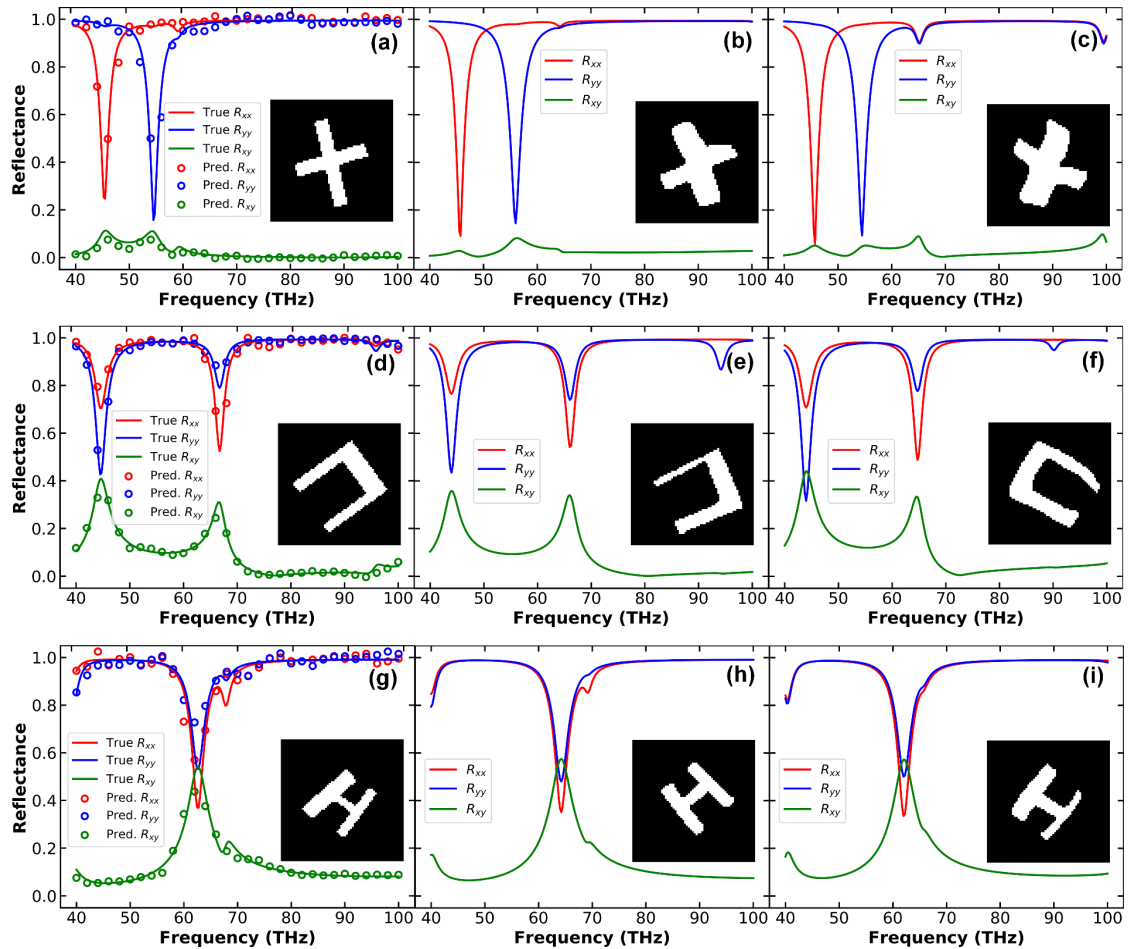

Figure 3. Evaluation of the deep generative model. (a), (d) and (g) are the samples from test dataset covering three geometry groups, where the solid lines and insets are the ground-truth reflection spectra and design pattern. The hollow circles are the output from prediction model, i.e., the predicted spectra discretized into 31 data points each. (b), (c), (e), (f), (h) and (i) are the retrieved designs and optical responses from the corresponding ground-truth reflection spectra.

For the inverse retrieval of the designs, we pick the 31 equidistant data points from the solid lines of each numerically simulated spectra in Figure 3(a), 3(d) and 3(g), and the concatenated 93-dimensional vectors act as one input to the generation model. Another input is the 20-dimensional latent variable, which is sampled from the prior standard Gaussian distribution of the latent space. The generation model accepts these two inputs, and produces the binary image representing meta-atom design. Instead of using only the required spectra as input, the stochastically sampled latent variable allows diverse output from the generation model, making it possible to solve the one-to-many mapping problem. Since the true latent distribution is different from the approximated Gaussian prior, as described by the KL-divergence in Equation 1, the sampled latent variable is not able to produce highly accurate retrieval from the given spectra in all cases. However, thanks to the prediction capability, we can feed the generated image back to the prediction model and compare the predicted spectra with the required spectra. Through this self-checking procedure, those highly precise retrieval results can be obtained by simply applying a threshold on the difference between forward prediction and requirements. Under a sum-square-error threshold of 0.2, we give two retrieval results demonstrated in Figure 3 (b), (c), (e), (f), (h) and (i). The insets are the retrieved design patterns, and the corresponding reflection spectra are obtained by numerically simulating the retrieved design. For the split ring and h-shape, the two model-retrieved geometries appear to be much similar to the ground-truth pattern with some symmetric transformations, while the retrieved cross patterns have larger arm width compared with the ground-truth shape. Nevertheless, all the generated designs reproduce the corresponding input reflection spectra with high fidelity, indicating that our model can effectively link the meta-atom design and optical response through the probabilistic representation by latent variables.

**On-demand inverse design of metamaterials**

Although our model performance is quite satisfactory on the test data set in both forward prediction and inverse retrieval, the real-world application usually requires on-demand inverse design of metamaterial with artificially defined optical response according to specific tasks. Based on the nature of electromagnetic resonance, we choose to describe each reflection spectrum as the sum of multiple Lorentz line shapes. Therefore, for each on-demand requirement on optical response, we only need to specify the resonant frequency $\omega_0$, bandwidth $\Delta$ and amplitude $A$ for one single resonant feature approximated by Lorentz line shape. The complete reflection spectrum for one set of input-output polarization configuration is given by summing up all the component resonances, as described below:

$$R_{xx,yy} = \sum_i \left(1 - \frac{A_i}{1+[(\omega-\omega_{0,i})/(\Delta_i/2)]^2}\right) \qquad (2)$$

The on-demand retrieved results are shown in Figure 4 for three cases with four generated meta-atom designs in each case. Similar to the previous model evaluation, the required spectra are fed to the generation model together with the sampled latent space from Gaussian prior, and the raw retrievals are filtered by a threshold of the difference between required spectra and the output from prediction model in order to guarantee accuracy. In Figure 4 (a), we assume a single resonance for both $R_{xx}$ and $R_{yy}$ with the same bandwidth of 4THz and amplitude A=0.7, at the frequency of 60 THz and 80 THz respectively. $R_{xy}$ is set as a constant of 0.1 across the frequency range. We demonstrate four out of the many retrieved meta-atom designs in the insets of Figure 4 (a), and the corresponding reflection spectra obtained by numerical simulations are plotted together as solid lines. Despite the very different meta-atom geometries, the optical responses of these retrieved designs agree fairly well with the required reflections. Similar results are also acquired in two other cases. In Figure 4(b), we require only the x polarization condition $R_{xx}$ has a single strong resonance at 70 THz with bandwidth of 4 THz and resonant dip of 0.1, while other two reflection spectra are set as constant of 0.1 or 1. In Figure 4(c), only $R_{yy}$ is required to have a dual band resonance with $A_1 = 0.7$, $\omega_{0,1} = 60$ THz, $\Delta_1 = 4$ THz, $A_2 = 0.7$, $\omega_{0,2} = 80$ THz, $\Delta_2 = 4$ THz, leaving the rest two spectra as constant of 0.1 or 1. In all cases, as long as the prescribed requirements are realizable for the sandwiched structure of the meta-atom, the proposed deep generative model can stably make diverse retrievals that best approximate the requirements.

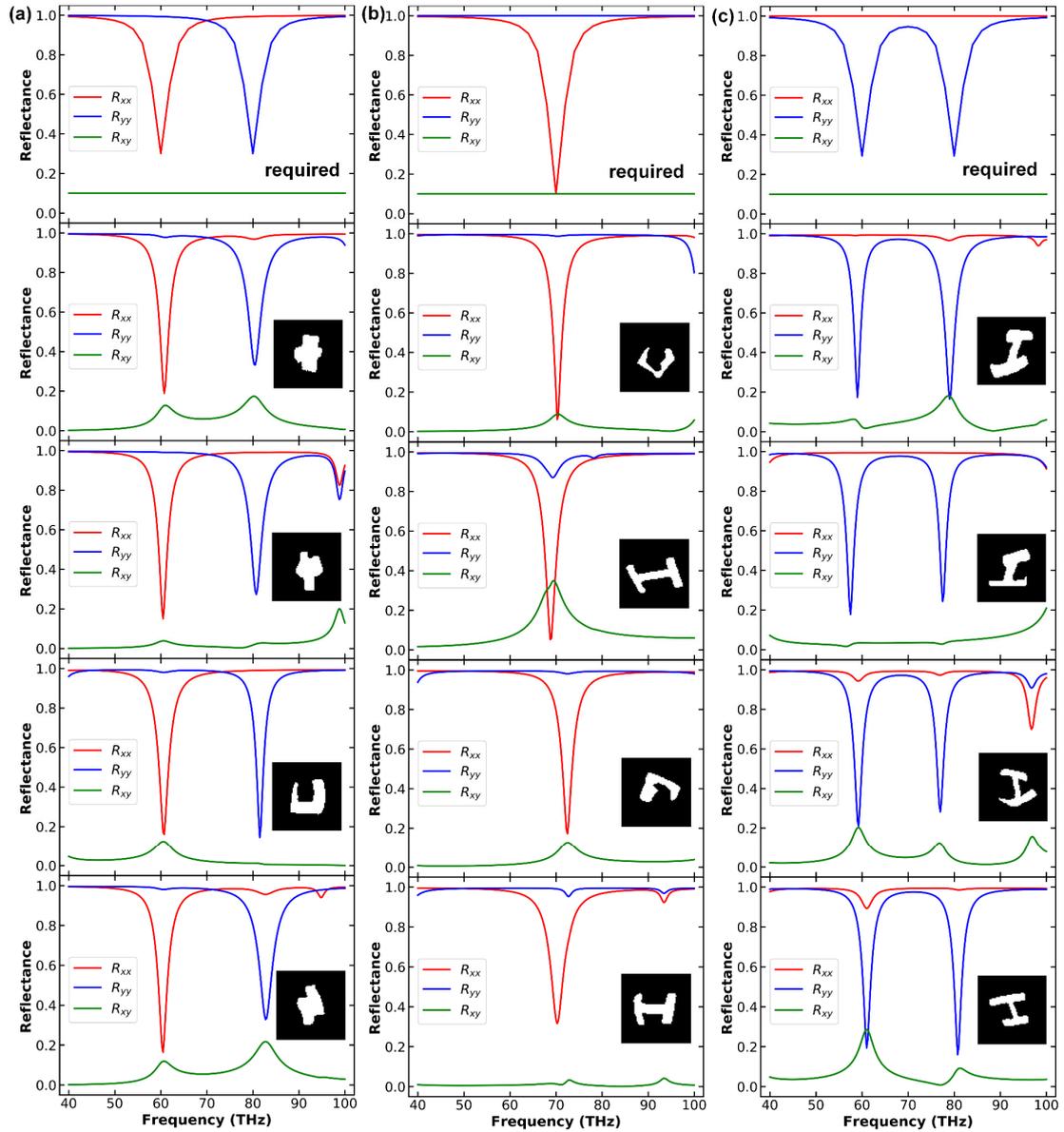

Figure 4: On-demand inverse design of metamaterials given specific reflection spectra. (a) Single resonance for $R_{xx}$ at 60 THz and $R_{yy}$ at 80 THz, (b) Only one resonance for $R_{xx}$ at 70 THz (c) Dual resonances for $R_{yy}$ at 60 THz and 80 THz. The topmost plot in each figure is the required reflection spectra specified by Lorentz line shape, and the rest four plots are examples of the retrieved meta-atom design (insets) and their corresponding reflection spectra from numerical simulation.

Also notably, albeit with optical responses highly matching the requirements, some of the retrieved designs have irregular geometries that appear not to belong to any of the geometry groups of cross, split ring or h-shape from our training set. This phenomenon indicates that by probabilistically representing the metamaterial in latent space, our deep generative model learns a more general distribution of the meta-atom design, which can help us to discover novel structures beyond previous experience. To further explore the structure of the latent space and its impact on inverse design, we deliberately sample the latent variables from different parts of the latent space to reconstruct the geometry. The required spectra are given in Figure 5 (a), where $R_{xx}$ and $R_{yy}$ have the same dual-band resonance line shape at 60 THz and 80 THz respectively with reflection dips of 0.5 and bandwidth of 4 THz, while $R_{xy}$ is also a dual-band Lorentz line shape with inverse resonant features. To make the inverse retrieval from different parts of the latent space, we first calculate the mean vectors of all the posterior Gaussian distributions of training data from the recognition model, and average the mean vectors within each geometry group to obtain the 20-dimensional center vector for different geometry groups, which can be viewed as the average vector representing the general designs of cross, split ring and h-shape respectively. Then, instead of directly sampling from the prior standard Gaussian distribution, we sample around the center vector of each geometry groups with a small standard deviation of 0.1. The retrieved designs and the corresponding reflection spectra are given in Figure 5(b) – 5(d), where the generated patterns have the geometry of cross, split ring and h-shape respectively, with good fidelity in the reproduced optical responses. Moreover, we can even sample around the midpoint of two different geometries. In consequence, the generated patterns will resemble both the seed geometries, such as a partial-cross-partial-h-shape in Figure 5(e) and a partial-cross-partial-split-ring in Figure 5(f).

Thanks to the well-learned latent space, we can make interpretable inverse retrieval according to specific preferences. This function is beneficial in applications with constraints on metamaterial geometry, such as in detectors, the metamaterial geometries need to be compatible with the mechanical etching hole structures [61]. Therefore, the latent variables not only offer an interpretable probabilistic representation of metamaterials, but also render the inverse design more flexible and versatile.

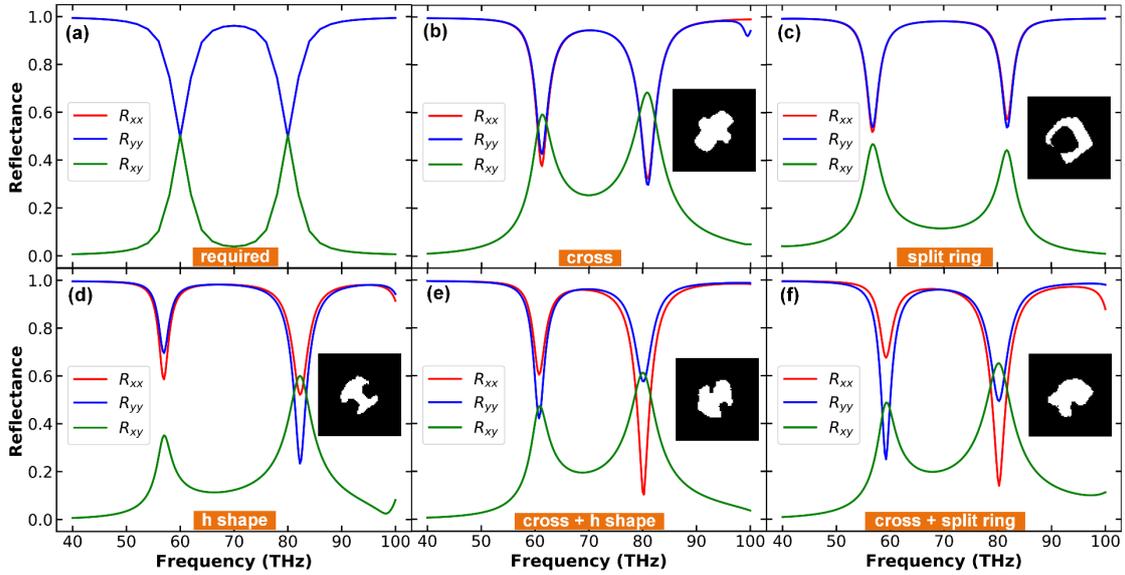

Figure 5: One-demand inverse retrieval of metamaterial design with latent variable sampled from given part of the latent space. (a) The required reflectance spectra. The retrieval results using a latent variable sampled around the center of cross cluster (b), split ring cluster (c) and h-shape cluster (d) in the latent space. The retrieval results using a latent variable sampled around the midpoint of cross cluster and h-shape cluster (e), and the midpoint of cross cluster and split ring cluster (f) in the latent space.

## Conclusions

To conclude, we propose and develop a deep generative model that elegantly solves the metamaterial design problem. Instead of building up a one-to-one mapping between meta-atom geometry and optical response that contradicts physical intuition, our model contains an encoder-decoder structure that first encodes the metamaterial design into a latent space, from which the latent variables are sampled as probabilistic representation of the metamaterial. On top of the latent variable, we can easily realize diverse retrieval of the meta-atom geometry given the required spectra, solving the one-to-many mapping problem in the inverse design. The forward prediction of optical response is achieved by a deterministic prediction model, which shares the extracted features with the encoder. Moreover, the encoder-decoder configuration in our model allows to utilize both labeled and unlabeled data in a semi-supervised learning scheme, which helps to exploit human experience to maximum extent economically without too much dependence on numerical simulations. The high efficiency and versatility of the proposed deep generative model make it a promising candidate in the research field of metamaterials and photonics, where the design is mainly based on physics-inspired methods in conjunction with numerical simulations from trial and error. The latent variable encoding and the subsequent generative process in our model render the inverse design more stable, diverse and versatile, and can be readily extended to other research domains of photonic and material science, enabling on-demand design, characterization and even new discoveries in various applications.

## Methods

In order to demonstrate the performance of the proposed deep generative model, we collected metamaterial patterns from three basic geometry groups: cross shape, split ring and h-shape. These three geometry groups were selected because they have been frequently adopted by previous works as efficient and fabrication-feasible candidates of meta-atoms and thus represent typical human experience in metamaterial design [11, 14, 17]. To increase diversity, we randomly sampled all possible design parameters such as length, width, relative offset and rotational angle of different parts of the geometries from the three groups, then applied random global distortion to the pattern before discretized to 64×64 binary image. Therefore, the generated metamaterial pattern covers a vast variety in design which can hardly be described by a few fixed parameters (some samples from the training data are given in Figure S3 of the Supplementary Material).

We collected 50000 patterns from each geometry groups, where we numerically calculated the optical response on 10000 of them. Hence, 20% of the training data were labeled as pattern-spectra pairs for deterministic learning, while the rest patterns only contributed to generative learning. We also prepared 1000 data for each geometry groups as test set. In the numerical simulation, CST Microwave Studio was employed to generate the reflection spectra data. The spacer was modeled as a lossless dielectric with permittivity of 2, and gold was treated by Drude model[62]. The proposed model was constructed under the open-source machine learning framework of *TensorFlow*.

## Acknowledgements

We acknowledge the financial support from the Office of Naval Research (N00014-16-1-2409).